\documentclass[a4paper,11pt]{amsart}
\usepackage{multicol,ifthen}
\usepackage{amssymb,stmaryrd}
\usepackage{harvard}
\usepackage{graphicx}

\newcommand{\R}{\mathbb R}

\newcommand{\C}{\mathbb C}

\newcommand{\beq}{\begin{equation}}
\newcommand{\eeq}{\end{equation}}
\newcommand{\beqarr}{\begin{eqnarray}}
\newcommand{\eeqarr}{\end{eqnarray}}
\newcommand{\beqa}{\begin{eqnarray*}}
\newcommand{\eeqa}{\end{eqnarray*}}
\begin{document}

\title[Cosmological spacetimes and  Weyl geometric scalar field]{Cosmological spacetimes balanced by a Weyl geometric scale covariant scalar field   }
\author[E. Scholz]{Erhard Scholz$\,^1$}\footnotetext[1]{\, scholz@math.uni-wuppertal.de\\
\hspace*{7mm}University Wuppertal, Department C, Mathematics and Natural Sciences,\\
\hspace*{6.5mm} and Interdisciplinary Center for Science and Technology Studies} 

\date{2008/11/21}

\begin{abstract}
A   Weyl geometric approach to cosmology is explored, with a  scalar field $\phi$ of (scale) weight $-1$ as  crucial ingredient   besides classical matter.
Its relation  to  Jordan-Brans-Dicke theory is analyzed; overlap and differences are discussed.  
 The energy-stress tensor of the basic state of the scalar field consists of a 
vacuum-like term $ \Lambda g_{\mu \nu }$  with $ \Lambda $ depending on the Weylian scale connection and, indirectly, on matter density. For a particularly simple class of Weyl geometric  models (called {\em Einstein-Weyl universes})  the
energy-stress tensor of the $\phi$-field can  keep space-time geometries in  equilibrium. A short glance at observational data, in particular supernovae Ia \cite{Riess_ea:2007}, shows encouraging empirical properties  of  these  models.
\end{abstract}

\maketitle 
\thispagestyle{empty}
\section{Introduction}
 For more than half a century,  Friedman-Lemaitre (F-L) spacetimes   have been serving as a successful paradigm  for  research in theoretical and observational cosmology. With  the specification of the parameters inside this model class,  $\Omega_m \approx 0.25,  \; \Omega_{\Lambda }\approx 0.75 $, new questions arise. Most striking among them are the questions of how  to understand the ensuing ``accelerated expansion'' of the universe indicated by this paradigm after evaluating the observational data,  and those concerning the strange behaviour of ``vacuum energy'' \cite{Carroll:Constant}. The latter seems to dominate the dynamics of spacetime  and of matter  in cosmically large regions, without itself being acted upon by the matter content of the universe. Such questions raise  doubts with respect to the reality claim raised by the standard approach \cite{Fahr:BindingEnergy}. They make it worthwhile to study to what extent small modifications in the geometric and dynamical presuppositions lead to different answers to these questions, or even to a different overall picture of the questions themselves.

In this investigation we study which changes of perspective may occur if one introduces scale covariance in the sense of integrable Weyl geometry (IWG) into the consideration of cosmological physics and geometry. At first glance this may appear as a  formal exercise (which it is to a certain degree), but the underlying intention is at least as physical as it is mathematical. The introduction of scale freedom into the basic  equations of  cosmology   stands in agreement with a kindred Weyl geometric approach to
scale covariant  field theory \cite{Drechsler/Tann,Drechsler:Higgs,Hung_Cheng:1988} and overlaps partially with conformal studies of   semi-Riemannian  scalar-tensor theories \cite{Fujii/Maeda:2003,Faraoni:2004}.
In distinction to the latter, Weylian geometry allows a mathematical overarching approach to cosmological redshift, without an 
ex-ante decision between the two causal hypotheses of its origin, space expansion or a field theoretic energy loss of photons over cosmic distances. 
Although  in most papers the first alternative is considered as authoritative, 
it is  quite interesting to see how in our frame the old hypothesis of a field theoretic reduction of photon energy, with respect to a family of cosmological observers, finds a striking mathematical characterization in Weyl geometry. Cosmological redshift  may here be expressed by the scale connection of Weylian Robertson-Walker metrics, in a   specific scale gauge (warp gauge). 

The following paper gives a short introduction to  basics of Weyl geometry  and the applied conventions and notations (section 2). After this  preparation the  scale invariant Lagrangian studied here  is introduced  (section 3). Different to Weyl's fourth order Lagrangian for the metric, an  adaptation of the standard Hilbert-Einstein action serves as the basis of our approach,  coupled in such a way  to a scale covariant scalar field $\phi$ of weight $-1$  that scale invariance of the whole term is achieved.  This approach is  taken over from W. Drechsler's  and H. Tann's research in  field theory, which explores an intriguing  path towards deriving mass coefficients for the electroweak bosons by coupling to gravity and the scalar field. For gravity theory Weyl geometric scalar fields show   similarities with  conformal studies of Brans-Dicke type theories, but differ  in their geometrical and scale invariance properties   (section 4).
Section 5 of this paper gives  a short outline  and commentary of Drechsler/Tann's proposal of the Weyl geometric Higgs-like ``mechanism''.  

In the  next section the variational equations of the Lagrangian are presented. They lead to a scale co/invariant form of the Einstein equations, a Klein-Gordon equation  for the scalar field and the  Euler equation  of ideal  fluids  (section 6).  
Then we turn towards cosmological modeling in the frame of Weyl geometry. The isotropy and homogeneity conditions of Robertson-Walker metrics are adapted to this context and lead to scale covariant Robertson-Walker fluids. New interesting features arise in the Weyl geometric perspective, in particular  with respect to the symbolic representation of cosmological redshift by a scale connection ({\em Hubble connection})  (section 7). The most simple Weyl geometric models of cosmology ({\em Weyl universes}) are similar to the classical static geometries; but here they are  endowed with a scale connection  encoding cosmological   redshift (section 8). 

Luckily,  the geometry of Weyl universes is  simple enough to allow an explicit calculation of the  energy-stress tensor of the scalar field's ground state (section 9). Thus it can be investigated under which conditions the scalar field safeguards dynamical consistency (equilibrium) of Weyl universes.  Those with positive space sectional curvature  are called {\em Einstein-Weyl universes}. A consistency condition derived from the Klein-Gordon equation of the scalar field leads to  specific coupling condition for   Hubble redshift to sectional curvature of the spatial fibres and thus to   the matter content of the universe.  

The article is rounded off by a short look at data from observational cosmology (section 10) and a  discussion of the perspective for cosmology opened up by the Weyl geometric approach to gravity and of some of the open questions (section 11).

\section{Geometric preliminaries and notations \label{preliminaries}}

We work in a classical spacetime given by  a differentiable  manifold $M$ of dimension $n=4$, endowed with a {\em Weylian metric}. The latter may be given by  an equivalence class $[(g,\varphi)]$ of pairs $(g, \varphi)$ of a Lorentzian metric $g= (g_{\mu \nu })$  of  signature $(-,+,+,+,)$,  called the {\em Riemannian component} of the metric,  and a {\em scale connection} given by a differential 1-form $\varphi= (\varphi_\mu) $.  Choosing a representative $(g,\varphi)$ means to  {\em gauge}  the metric. A {\em scale  gauge transformation}  is achieved by rescaling the Riemannian component of the metric and an associated transformation of the scale connection
\beq  \tilde{g} =  \Omega ^2 g \; , \quad \quad  \tilde{\varphi} = \varphi - d \log \Omega \, , \label{gauge transformation}\eeq 
where $\Omega > 0$ is a strictly positive real function on $M$.   

Einstein's famous argument against Weyl's original version of scale gauge geometry (stability of atomic spectra) and --- related to it ---  coherence with quantum physics  \cite{Audretsch_ea} make it advisable, to say the least, to   restrict the Weylian metric to one with  {\em integrable scale (``length'') connection}, 
$  d \varphi = 0 $.

 The  integration of $\varphi$ leads to a  {\em scale} (or ``length'') {\em  transfer} function $\lambda(p_o,p_1)$  allowing to compare metrical quantities at different points of the manifold,
 \beq \lambda (p_o,p_1) = e^{\int_{u_0}^{u_1}  \varphi(\gamma'(u))du}\, , \label{scale transfer}\eeq
 $\gamma(u)$ any differentiable  path  from a fixed reference point $p_0=\gamma(u_0)$  to $p=\gamma(u_1)$. 
 In simply connected regions the 
scale connection can be integrated away, 
 $\tilde{g} = \lambda ^2  g$,  $\tilde{\varphi}= 0$, if $d\varphi=0$.  
 In this case the  
Weylian metric may be written in Riemannian form, but need not. 
By obvious reasons this gauge is called {\em Riemann gauge}.
Thus  one may  work in {\em integrable Weyl geometry (IWG)}  {\em without passing to Riemann gauge} by default.

 There is a uniquely determined {\em Levi Civita connection} of the Weylian metric,  
\beq  \label{Christoffel}  \Gamma^{\mu }_{\nu \lambda } =   {}_g\Gamma^\mu _{\nu \lambda } + \delta ^{\mu }_{\nu } \varphi _{\lambda } +
\delta ^{\mu }_{\lambda } \varphi _{\nu } - g_{\nu \lambda } \varphi^{\mu } \, . \eeq 
Here  $ {}_g\Gamma^\mu _{\nu \lambda }$ denote the coefficients of the affine connection with respect to the Riemannian component $g$ only. The Weyl geometric covariant derivative with respect to $\Gamma^{\mu }_{\nu \lambda }$ will be denoted by $\nabla_ {\mu } $; the covariant derivative with respect to the Riemannian component of the metric only by ${}_g\hspace{-0.1em}\nabla_ {\mu } $.  $\nabla_ {\mu } $ is  an invariant operation for  vector and tensor fields on $M$, which are themselves  invariant under gauge transformations. The same can be said for {\em geodesics} $\gamma_W$ of Weylian geometry, defined by $\nabla_ {\mu } $,  and for the  {\em curvature tensor} $R = R^{\alpha} _{\beta \gamma \delta}$ and its contraction, the Ricci tensor $Ric$. All these are {\em invariant} under scale  transformations.  

For calculating geometric quantities (covariant derivatives, curvatures etc.) of a  Weylian 
metric in the gauge  $(g,\varphi)$ one may   start from the corresponding (Riemannian) ones,  with respect to the Riemannian component $g$ of the Weylian metric given by $(g,\varphi)$. Like for the affine connection we use the 
pre-subscript $g$ to denote the latter, e.g., ${}_g\overline{R}$ for the scalar curvature of the Riemannian component. For $dim \,M = n$ we know already from \cite[p. 21]{Weyl:InfGeo}: 
\beqarr  \quad \quad \quad \quad  \overline{R} &=& {}_g\overline{R} - (n-1)(n-2) \,\varphi_{\lambda }\varphi^{\lambda } - 2(n-1) \, {}_g\hspace{-0.1em}\nabla_{\lambda }\, \varphi^{\lambda }  \label{Weylian scalar curvature} \\
(Ric)_{\mu \nu } \hspace{-0.5em} &=& {}_gRic_{\mu \nu } + (n-2) ( \varphi_{\mu }\varphi_{\nu } -  {}_g\hspace{-0.1em}\nabla_{( \mu } \varphi_{\nu )} )  \label{Weylian Ricci curvature} \\
& & \hspace*{7mm} - g_{\mu \nu }((n-2) \varphi_{\lambda }\varphi^{\lambda } + {}_g\hspace{-0.1em}\nabla_{\lambda }\varphi^{\lambda} )\, \nonumber
\eeqarr
For $n=4$, in particular, that is 
\[  \overline{R} = {}_g\overline{R} - 6(  \,\varphi_{\lambda }\varphi^{\lambda } + \, {}_g\hspace{-0.1em}\nabla_{\lambda }\, \varphi^{\lambda } ) \quad etc.\]

In order to make full use of the  Weylian structure on $M$ one often considers (real, complex etc.)  functions $f$ or (vector, tensor, spinor \ldots) fields  $F$ on $M$, which transform under gauge transformations like
\[  f \longmapsto \tilde{f} = \Omega ^k f \;  ,\quad \quad F \longmapsto \tilde{F} = \Omega ^l F  \; . \]
$k$ and  $l$ are the (scale or Weyl) {\em weights} of $f$ respectively $F$. We write 
$  w(f):= k \; , \quad w(F) := l  $ and speak of {\em Weyl functions} or {\em Weyl fields} on $M$. To be more precise mathematically, Weyl functions and Weyl fields are equivalence classes of ordinary (scale invariant) functions and fields. Obvious examples are: $w(g_{\mu \nu })=2$, $w(g^{\mu \nu })=-2$ etc.  As the curvature tensor $R = R^{\alpha} _{\beta \gamma \delta}$  of the Weylian metric and  the Ricci curvature tensor $Ric$ are scale invariant,   scalar curvature $  \overline{R} = g^{\alpha \beta } Ric\, _{\alpha \beta } $
is  of weight $ w(\overline{R}) =-2$.

Formulas similar to (\ref{Christoffel}) to (\ref{Weylian Ricci curvature}) are derived for  conformal transformations in 
semi-Riemannian gravity \cite[chap. 3]{Fujii/Maeda:2003}, \cite[chap. 1.11]{Faraoni:2004}. But there the  geometrical and physical meaning is slightly different. While in semi-Riemannian  relativity these equations  are used  to  calculate  the affine connection and curvature quantities of an ``original'' metric $g$ after a conformal mapping to a different one,  $g_{\ast}= \Omega^2 g$,  Weyl geometry considers conformal rescaling as a {\em gauge transformation in the original sense of the word}, expressing the change of  measuring devices (or equivalently of dilatations).  The  aim her is to study  scale {\em covariant} behaviour of quantities and structures, with particular attentiveness to scale {\em invariant} aspects.

Note  that the application of Weyl's  covariant derivative $\nabla$, associated to the Weyl geometric affine connection (\ref{Christoffel}),  to  Weyl fields $F$ of weight $w(F) \neq 0$   {\em does not} lead to a  scale covariant quantity. This deficiency can be repaired by introducing a {\em scale covariant derivative} $D_{\mu }$ of Weyl  fields in addition to the scale invariant  $\nabla _{\mu }$ \cite{Dirac:1973}, \cite[app.  A]{Drechsler/Tann}):
\beq D F := \nabla F  + w(F)
\varphi \otimes F \,.    \label{scale covariant derivative}  \eeq
Thus, for example, a scale covariant vector field $F^{\nu }$ has the scale covariant derivative 
\[  D_{\mu } F^{\nu } := \partial _{\mu } F^{\nu } + \Gamma^{\nu }_{\mu \lambda } F ^{\lambda} + w(F)\, \varphi_{\mu } F^{\nu }  \, , \]
with the abbreviation
$\partial_{\mu } := \frac{ \partial }{\partial x^{\mu } }$ etc. 

For the description of relativistic trajectories Dirac  introduced {\em scale covariant} geodesics $\gamma $ with weight of the tangent field $u:= \gamma '$  $w(u)=-1$, 
 defined by
\beq D_{u} u = 0
 \eeq 
with scale covariant derivation $D$, i.e.
\[  \nabla_{\mu} \gamma '^{\mu}  \gamma '^{\nu} - \varphi_{\mu}\gamma '^{\mu}  \gamma '^{\nu} = 0 \quad \mbox{for } \; \nu = 0, \ldots n-1 \; . \]
They differ from Weyl's scale invariant geodesics $\gamma _W$ only by parametrization. 

By construction  $g(u,u)$ is scale invariant. In particular $g(u,u)= \pm 1$ for spacelike or  timelike geodesics; and {\em  geodesic distance} $d_{(g,\varphi)}(p_o,p_1)$ between two points $p_o, p_1$ with respect to $g$  is given by the parameter  of the Diracian geodesics. Of course it  depends on the scale gauge $(g, \varphi)$ chosen and coincides with (semi-Riemannian) distance of  {\em Weyl geometric geodesics} measured in the {\em Riemannian component} of $(g,\varphi)$
\beq d_{(g,\varphi)}(p_o,p_1) = \int _{\tau_o}^{\tau_1} \sqrt{|g(\gamma ', \gamma ')|} \, d\tau =  (\tau _1 - \tau _o)\, . \label{geodesic distance}
\eeq

 Dirac's scale covariant geodesics  have the same scale weight as energy $E$ and mass $m$, $w(E) = w(m)= -1$, which are  postulated  in order to keep  scaling  consistent with the Planck relation $E= h \nu$ and Einstein's $E = mc^2$ (with true constants $h$ and $c$).
 Thus mass or energy factors assigned to particles or field quanta can be described  in a gauge independent manner in Dirac's calculus: one just has to associate {\em constant mass factors} $\hat{m}$ (more formally defined  below) to the Diracian geodesics; the scale gauge dependence is implemented already in the the latter.

For any nowhere vanishing Weyl function $f$ on $M$ with weight $k$ there is a gauge (unique up to a constant), in which $\tilde{f}$ is constant. It is given by (\ref{gauge transformation}) with
\beq \Omega = f^{-\frac{1}{k}} \;  \label{f gauge}
\eeq
 and will be called   {\em f-gauge} of the Weylian metric. There are infinitely many gauges, some of which may be  of mathematical importance in specific contexts.
An 
$\overline{R}$-gauge (in which scalar curvature is scaled to a constant) exists for manifolds with nowhere vanishing scalar curvature. It ought to be called   {\em Weyl gauge}, because Weyl assigned a particularly important role to it in his foundational thoughts about  matter and  geometry  \cite[298f.]{Weyl:RZM5}. 

Similar to  ordinary semi-Riemannian and conformal scalar tensor  theories  of gravity,  one often considers  a {\em nowhere vanishing scalar field $\phi$} of weight $w(\phi)= -1 $  in Weyl geometry. Originally the name-giving authors of   
Jordan-Brans-Dicke (J-B-D) theory hoped to find  a ``time varying'' real scalar field as an empirically meaningful device, corresponding to the so-called   {\em Jordan frame} of 
J-B-D theories  \cite{Brans:2004}. But  empirical evidence, gathered  since the 1960-s, and a theoretical reconsideration of the whole field since the 1980-s have accumulated overwhelming arguments that  the conformal picture of the theory with $\phi$ scaled to constant, the so-called  {\em Einstein frame},  if any at all,  ought to be considered as  ``physical'', i.e. of empirical content  \cite{Faraoni_ea:Transformations}. 

A similar view holds in the Weyl geometric approach. There the norm $|\phi|$ of the scalar field (now complex or even a Higgs-like isospinor complex two-component field) may be considered as setting the physical scale \cite{Drechsler/Tann,Drechsler:Higgs}.
That leads to an obvious method to extract the {\em scale invariant magnitude} $\hat{X}$ of a scale covariant {\em local quantity} $X$ of weight $w:= w(X)$, given at point $p$. One just has  to consider the  proportion with the appropriately weighted power of $|\phi|$. In this sense the {\em observable magnitude}  $\hat{X}(p)$ of a Weyl field $X$ with $w=w(X)$ is given by
\beq
  \hat{X}(p) := \frac{X(p)}{|\phi|^{-w}(p)} = X(p)\, |\phi|^w(p) \; .  \label{observable}
\eeq
By definition $\hat{X}$ is {\em scale invariant}. 

For example the scale invariant length $\hat{l}(\xi)$ of  a vector  $\xi$ at $p$, $\xi \in T_pM$, is  $\hat{l}(\xi) =|\phi|\, | g(\xi,\xi)|^{\frac{1}{2}} $,  independent of the scale gauge considered. Matter energy density in the sense of $\rho = T^{(mat)}_{00}$ (cf. equ. (\ref{classicalmatter tensor})) has to be compared with observed quantities by  $\hat{\rho }= \rho |\phi|^2$, etc..
  Geodesic distance in the sense of  (\ref{geodesic distance})  is a  
non-local, scale dependent concept; but its observable $\hat{d}(p_o.p_1)$, i.e.  the  {\em scale invariant distance} between two points,  is calculated by integrating local ``observables'' derived from the infinitesimal arc elements. For geodesics $ds = |g(\gamma ',\gamma ') |^{\frac{1}{2}}=1$ and thus
\beq   \hat{d}(p_o,p_1)  =   \int _{\tau_o}^{\tau_1} |\phi| \, d\tau    \;  . \label{invariant distance}\eeq
This is identical to geodesic distance in $|\phi|$-gauge.

Choosing  the scale gauge such that the norm of $\phi$ becomes constant may thus facilitate the  calculation of  scale invariant observables considerably. In $| \phi |$-gauge $\hat{X}$ is identical to $X$ up to a (``global'') constant factor depending on measuring units.
 If lower $\ast$ denotes values in $|\phi|$-gauge, we clearly have 
\beq \hat{X} \doteq const \,  X_{\ast}\, \quad \mbox{with} \quad const =  |\phi_{\ast}|^{w(X)} \, . \;\;  \label{X phi gauge}
\eeq  
The dotted  $\doteq$ indicates that the equality  only holds in a specified gauge (respectively frame, if one considers the analogous situation in conformal scalar tensor theories, cf. section 4).
 In this sense $|\phi|$ gauge is  physically {\em  preferred}. Observables $\hat{X}$ are directly read off from the quantities given in this gauge, $\hat{X} \sim X_{\ast}$. In particular for distances
 \[ \hat{d}(p_o,p_1) \doteq d_{\ast}(p_o,p_1) \, , \]
 up to a global constant.  If scale invariant local quantities of Weyl geometric gravity (with a scalar field) are empirically meaningful,
 $|\phi|$-gauge expresses directly the behaviour of atomic clocks or other physically distinguished measuring devices. On the other hand, there may be mathematical or other reasons to calculate $\tilde{X}$ in another scale gauge first.

The physical fruitfulness of Weylian geometry (in the scalar field approach) depends on the answer to the following question:  
Can  measurement by atomic clocks be characterized by scale invariant classical observables  like above? --- Those who stick to the default answer that this is not the case and Riemann gauge expresses observables directly will be led back to Einstein's semi-Riemannian   theory.  If this were the only possibility,  the generalization to IWG would be  redundant.  However, this is not at all the case when we consider  a Weyl geometric version of scalar tensor gravity with the  assumption that $|\phi |$ ``sets the scale'' (in the sense above).

\section{Lagrangian }

We start from  scale invariant Lagrangians similar to those studied in conformal
 J-B-D type theories of gravity \cite{Fujii/Maeda:2003,Faraoni:2004} with a real scalar field $\phi$. \citeasnoun{Tann:Diss} and \citeasnoun{Drechsler/Tann} have investigated the properties of  a complex   version of it in their field theoretic studies of a Weyl geometric unification of gravity with electromagnetism.   \citeasnoun{Drechsler:Higgs}  even includes 
 semi-classical  fields of the standard model (fermionic and bosonic), extending $\phi$ to  a Higgs-like  isospinor spin 0 doublet.  Here we deal exclusively with gravity and  might specialize to a real scalar field, but we do not. 

In order to indicate the symbolical interface to the extension of the Weyl geometric approach to   electromagnetism and/or the standard model sector of elementary particle physics (EP), studied by Drechsler and Tann, we stick (formally) to a complex version of the scalar field $\phi$, although for our purposes we are essentially concerned with $|\phi|$ only. The Lagrangian is 
\beq
 {\mathcal L}  =   \sqrt{|g|}  \left( L^{(HE)} +  L^{(\phi )} +    
   \ldots (L^{EP} \ldots)   +  L^{(em)} +  L^{(m)}  \right)  
 \, , \label{total Lagrangian A} 
\eeq
where $ |g| = |det (g_{ij})|$. 
Standard model  field theoretic Lagrangian terms, $L^{(EP)}$,  are indicated in brackets (cf. section \ref{section extension}).
$\mathcal{L}^{(HE)} $ is the Hilbert-Einstein  action in scale invariant form due to coupling of the scalar curvature to a complex scalar field $\phi$,  $w(\phi)= -1$.  $\mathcal{L}^{(\phi)} $ is the  scale invariant Lagrangian of the scalar field,   $ \mathcal{L}^{(m)}$   the Lagrangian of classical matter for  an essentially phenomenological characterization of mean density  matter. Here we consider the most simple form of a neutral fluid (even dust). A  more sophisticated (general relativistic magnetohydrodynamical)   $ \mathcal{L}^{(m)}$ will be necessary for more refined studies, e.g.  of structure formation arising from hot intergalactic plasma of intergalactic jets etc.\footnote{For a first heuristic discussion of structure formation compatible with Einstein-Weyl models, cf. \cite[chap. 6]{Fischer:Balance}.}  
\beqarr
L^{(HE)} &=& \frac{1}{2} \xi \,(\phi^{\ast} \phi)^{\frac{n-2}{2}}  \overline{R}   \, , \quad \quad w(\phi)=-1, \nonumber \\
  L^{(\phi )} &=&  -\left( \frac{1}{2}  D^{\mu }\phi^{\ast}D_{\mu} \phi   -  V(\phi) \right) \label{Lagrangian phi} \\
 L^{(m)} &=&   \mu (1+ \epsilon ) \, , \quad \quad w(\mu)=-4, \; \nonumber \\
 L^{(em)} &=&  \frac{[c]}{16 \pi}F_{\mu \nu } F^{\mu \nu }   \nonumber
\eeqarr
$\xi = \frac{n-2}{4(n-1)}$   ($n=$ dimension of spacetime) is the known coupling constant establishing conformal invariance  of the action  $\mathcal{L}^{(HE)} + \mathcal{L}^{(\phi )}$, if   covariant differentiation and  scalar curvature refer to the {\em  Riemannian component of the metric only} ($_g \hspace{-0.1em} \nabla _{\mu }$ and $_g\overline{R}$ in the notation above) \cite{Penrose:Conformal,Tann:Diss}. Here $n=4, \xi = \frac{1}{6}$.

The potential term in  $\mathcal{L}^{(\phi )}$ is  formal placeholder for a quadratic mass like and a biquadratic self interaction term  
\beq V(\phi) = \lambda _2 (\phi^{\ast} \phi) + \lambda _4  (\phi^{\ast} \phi)^2 \; ,\quad  \;  \; w(\lambda_2 )=-2, \;  w(\lambda_4 )=0, \label{potential}
\eeq 
with scale invariant coupling constant  $\lambda _4$  like in \cite{Drechsler:Higgs} and scale {\em co}variant quadratic coefficient $\lambda _2 $. 
Formally, $V$ looks like the Lagrange term of a scale covariant cosmological ``constant''. We shall see, however (equ.  (\ref{Lambda})), that the energy stress tensor of $\phi$ contains other, more important contributions.

 Our {\em matter Lagrangian} consists of  a fluid term with energy density $\mu$ and   internal energy ratio $\epsilon $ similar to the one in  \cite[69f.]{Hawking/Ellis}, with functions $\mu , \varepsilon $  on spacetime of weight $w(\mu)= -4 $, $w(\varepsilon) = 0$.  
$L^{(m)}$ is related   to  {\em timelike unit vectorfields} $X = (X^{\mu })$ of weight $w(X) = -1$ representing the flow  and constrained by the condition that during variation of the flow lines  its energy density flow 
\[ j := \mu (1+ \epsilon ) \, X \,  \]
satisfies the local energy conservation of the matter current
\beq  div \, j  = D_{\mu  }  j^{\mu } = 0 \, . \label{fluid constraint}  \eeq
For abbreviation we set
\beq
\rho  := \mu (1+ \epsilon )\,  . \label{rho} 
 \eeq

As an alternative, one might try  to model classical matter by a second scalar field $\Phi$ with the same scale weight as $\phi$ \cite[chap. 3.3]{Fujii/Maeda:2003}. The coupling to the scalar field $\phi$ \cite[(3.60)ff.]{Fujii/Maeda:2003} could  be transformed into a Weyl geometric kinetic term $g^{\mu \nu}D_{\mu}\Phi D_{\nu}\Phi $ with scale covariant derivative  $D_{\mu}\Phi  = (\partial _{\mu} - \varphi_{\mu}) \Phi $. But for cosmological applications the (observational) restriction of negligible pressure, would lead to an  artificial coupling between matter and the scalar field $\phi$. We therefore choose here the approach adapted  from \cite{Hawking/Ellis}.

 In terms of extension of gauge groups, the Weyl geometric approach to gravity  works in the frame of  the scale extended Lorentz or
 Poincar\'e group, sometimes called the {\em (metrical) Weyl group} 
\[ W \cong \R^4 \ltimes SO(1,3) \times \R^{+} \, . \]
For inclusion of  standard model (EP) fields it has to be extended by    internal symmetries  $SU_3 \times SU_2 \times U(1)_Y$ .

To bring the constants in agreement with Einstein's theory,  the constant in 
$|\phi|$-gauge has to be chosen such that the coefficient of the Hilbert Einstein action becomes 
\beq   \frac{1}{2} \xi |\phi_{\ast}|^2 \doteq \frac{[c^4]}{16 \pi G}  \, . \label{G phi}
\eeq
  For $n=4$  that means 
\beq 
|\phi_{\ast}|^2 \doteq 6 \frac{c^4}{8 \pi G} \, ,
  \eeq 
In other words 
\beq   |\phi_{\ast}|  \sqrt{\hbar c} \doteq  \frac{1}{2} \sqrt{\frac{3}{\pi} E_{Pl}} \approx \frac{1}{2} E_{Pl}\, , \quad \quad  |\phi_{\ast}|^{-1}  \sqrt{\hbar c}  \doteq 2 \sqrt{\frac{\pi}{3}}  l_{Pl} \approx 2 l_{Pl} \, ,
\eeq
with $E_{Pl}$, $l_{Pl} $ Planck energy, respectively  Planck length. 

Some authors conjecture \cite{Hung_Cheng:1988,Smolin:1979,Hehl_ea:1989,Mielke_ea:dark_matter_halos} that a condensation, close to the Planck scale, of an underlying non-trivial scale bosonic  field $\varphi$ with $d \varphi \neq 0$ may give a deeper physical reason for the assumption that   $\phi$  ``sets the scale" in the sense of (\ref{observable}) and (\ref{G phi}). If this were  true, the scalar field $\phi$, and with it the integrable scale connection $\varphi$ taken into consideration here, would probably characterize a macroscopic state function of some kind  of scale boson condensate. This is an interesting thought, but  at present a  reality claim for this conjecture would be premature.

\section{$|\phi|$-gauge and scalar tensor theories of gravity \label{scalar tensor theories}}

In a formal sense, our Lagrangian may be considered as belonging to the wider family of scalar-tensor theories of gravity. The scale invariant Hilbert-Einstein action is  analogous to the one of 
Jordan-Brans-Dicke  theory. One should keep in mind, however, that the conceptual frame and the (model) dynamics are different. In  J-B-D theories  rescaling of the metric $g_{\ast}=\Omega ^2 g$ expresses a conformal mapping in which the affine connection and curvature quantities derived from $g$  are ``pulled back'' to the new  frame and expressed in terms of $g_{\ast}$. Two conformal pictures, usually called  ``frames'' ($g$ or $g_{\ast}$), represent possible different physical models \cite{Faraoni_ea:Transformations}. Thus the question arises which of the pictures (frames) may be  ``physical'', if any. \citeasnoun{Faraoni_ea:Transformations}
give  strong arguments in favor of the conformal picture in which the factor $|\phi|^2$ in the
Hilbert-Einstein action is normalized to a constant, the so-called {\em Einstein frame}. Some authors have  started to look for a bridge between scalar tensor theories and Weyl geometry \cite{Shojai:2000,Shojai/Shojai:2003}.

In the Weyl geometric approach all scale gauges are, in principle, equivalent. Weyl geometry is a scale invariant structure; local physical quantities $X$ (locally defined ``lengths'', energy densities, pressure, \ldots) are scale covariant (transform according to their gauge weight), but have scale invariant observable quantities  $\hat{X}$, cf.  (\ref{observable}). In this sense, Weyl geometry is a {\em gauge theory} like any other (needless to remind that it has given the name to the whole family). On the other hand,  scale invariant quantities  $\hat{X}$ can  be read off  directly in $|\phi|$-gauge up to a constant factor.
In this respect and different to other gauge theories, $|\phi|$-gauge provides us with a {\em preferred scale}. This corresponds well to the established knowledge that atomic clocks etc. define a physical scale, a fact which cannot be neglected in any reasonable theory of gravity, cf.
 \cite{Quiros:2000,Quiros:2008}.

In  integrable Weyl geometry, which we consider here exclusively (cf. section \ref{preliminaries}), the scale connection $\varphi$ is ``pure gauge'', i.e. has curvature zero, and can be integrated away. So   there are two distinguished gauges, one in which the scale connection is gauged away  and one in which the norm of the scalar field is trivialized, i.e. made constant. The complex, or isospin, phase of $\phi$ plays its  part only if electromagnetic fields or weak interaction is considered (is ``switched on''), cf. \cite{Drechsler/Tann,Drechsler:Higgs}; {\em here} we  abstract from its  dynamical role:
\beqa
(\tilde{g}, \tilde{\varphi})\, , &\tilde{\phi},&  \quad \mbox{with } \quad \tilde{\varphi}= 0 \,  \;   \, \quad  \quad \quad \mbox{({\em Riemann gauge}) } \;\; \\
(g_{\ast}, \varphi_{\ast}) \, , &\phi_{\ast}, & \quad  \mbox{with }  \,  \; |\phi_{\ast}| = const  \,  \quad \mbox{({\em $|\phi|$-gauge}) } \;\; 
\eeqa
 Formally the Einstein frame of J-B-D theories corresponds   to $|\phi|$-gauge,  Jordan frame (more precisely one of its choices) to Riemann gauge. 

Scale connection 
$\varphi_{\ast}$ (of $|\phi |$-gauge) and scalar field $|\tilde{\phi}|$ (in Riemann gauge) determine each other. Structurally speaking  they are different aspects of the same entity (in {\em integrable} Weyl geometry).
 The   scalar field  in Riemann gauge, more precisely its norm,  can be written as
\beq | \tilde{\phi}|  =  |\tilde{\phi}(p_o)| e^{- \sigma } \quad \mbox{with } \;\;
\sigma (p) = \int_{p_o} ^{p} \varphi_{\ast}  (\gamma ' ) \; , \label{sigma}
\eeq
$\gamma$ connecting path between $ p_o, \, p$, and $\varphi_{\ast}$ the scale connection in $|\phi|$-gauge.
It is just the inverse of Weyl's ``length'' (scale) transfer function (\ref{scale transfer}) in $|\phi|$-gauge up to a constant, $|\tilde{\phi}| \sim  \lambda(p_o,p)^{-1}$. The other way round, the scale connection $\varphi_{\ast}$ in $|\phi|$-gauge can be derived from the scalar field $\tilde{\phi}$ in Riemann gauge,
\beq  \varphi_{\ast} = d \sigma = - d \log |\tilde{\phi}|= - \frac{d |\tilde{\phi}|}{|\tilde{\phi}|}\, , \quad \mbox{i.e. }  \;\; \varphi_{\mu}= \partial _{\mu} \sigma \,.  \label{varphi_ast}
\eeq 

 The dynamics of $\varphi$ (in $|\phi|$-gauge) is governed by the Lagrangian of the $\phi$-field in (\ref{Lagrangian phi}),   $\mathcal{L}^{(\phi)} =   \sqrt{|g|} (-\frac{1}{2}D^{\mu }\phi^{\ast}D_{\mu} \phi   +  V(\phi) )  $. In IWG $\varphi$ cannot have a scale curvature term ``of its own'' ($d\varphi$ vanishes). This does {\em not} mean that the Weyl geometric extension of classical gravity is dynamically  trivial. In Riemann gauge its 
non-triviality is obvious.
At first glance it may appear  trivial  in $|\phi|$-gauge, because $|\phi |= const$. A second glance shows, however, that it is not, due to the scale connection terms of the covariant derivative. The dynamics of the scalar field in Riemann gauge is now expressed by a Lagrangian term in the scale connection $\varphi = d \sigma $, i.e., in the derivatives of $\sigma $.   

It may be useful to compare  the $|\phi|$-gauge Lagrangian 
\beq \mathcal{L}^{(EH,\, \phi)} 
\doteq  \sqrt{|g_{\ast}|} \left( \frac{1}{2}\xi\,  \overline{R_{\ast}} - \frac{1}{2} D^{\mu }\phi_{\ast}^{\ast}D_{\mu} \phi_{\ast}   +  V(\phi_{\ast})\right)   \label{phi gauge Lagrangian}
\eeq
with  the the corresponding expression of semi-Riemannian scalar tensor theory. If   non-gravitational (em or ew) interactions  are abstracted from, $|\phi|$ can be considered as an essentially  real field
\[ \phi = |\phi| \, . \] 

The corresponding expression in Einstein frame \cite[chap. 3.2]{Fujii/Maeda:2003}
\beq  
\sqrt{|g_{\ast}|} \left( \frac{1}{2}\xi \, _{g_{\ast}}\overline{R_{\ast}} - \frac{1}{2} (1+ 6\xi)\,  g_{\ast}^{\mu \nu} \partial_{\mu}  \sigma  \partial_{\nu}  \sigma  + V(\phi_{\ast}) \right)
\eeq
is variationally equivalent to (\ref{phi gauge Lagrangian}). It just 
has shifted the $6  \varphi_{\ast}^{\mu} \varphi_{\ast \mu} $ term of (\ref{Weylian scalar curvature}), plugged into (\ref{phi gauge Lagrangian}), to the kinetic term in  $\partial_{\mu} \sigma = \varphi_{\ast \mu}$.
 The last term, $6 _g\nabla_{\lambda } \varphi^{\lambda }$,  in  (\ref{Weylian scalar curvature}) is a gradient and has no consequence for the variational equations.\footnote{Fujii/Maeda's $\sigma$ contains a factor $\sqrt{1 + 6 \xi} $, compared with our's and a different sign convention for $V$ \cite[(3.28), (3.30)]{Fujii/Maeda:2003}.}

Nobody would consider semi-Riemannian scalar tensor theories in Einstein frame dynamically trivial. This  comparison may thus help to understand  that even a scale connection with $d \varphi = 0$ {\em can play  a dynamical role} in Weyl geometric gravity.  Below we shall study a  simple example, where  $\varphi$ even assumes {\em constant values} in large cosmological ``average'', respectively idealization. We have to keep  in mind that even then the Weylian scale connection $\varphi$ indicates a dynamical element of spacetime. ``Statics'' is  nothing but  a dynamical constellation in equilibrium.

\section{Extension to the field theoretic sector \label{section extension}}

Field theoretic contributions to the Lagrangian (electroweak, Yukawa, fermionic), adapted  from conformal field theory, are studied  in \cite{Drechsler:Higgs,Hung_Cheng:1988} and other works:
\beqarr
L^{(ew)} &=& \alpha _1 \left( W_{\mu \nu}  W^{\mu \nu}  +  B_{\mu \nu}  B^{\mu \nu} \right) \nonumber \\ 
L^{(Y)} &=&  \alpha _2\, \left( (\bar{\psi} _L \tilde{\phi})\Psi_R  +  (\bar{\psi} _R \tilde{\phi})\Psi_L \right)  \nonumber \\
 L^{(\Psi )} &=&    \frac{i}{2} (\bar{\Psi}_L \gamma ^{\mu} \tilde{D}_{\mu }\Psi_L - \bar{\Psi}_L  \tilde{D}_{\mu }\gamma ^{\mu}\Psi_L  ) +   \nonumber  \\
\hspace{20mm } & & \hspace*{40mm} + \frac{i}{2} (\bar{\Psi}_R \gamma ^{\mu} \tilde{D}_{\mu }\Psi_R - \bar{\Psi}_R  \tilde{D}_{\mu }\gamma ^{\mu}\Psi_R  ) \nonumber
\eeqarr
 $\Psi $ denotes left and right handed spinor fields of spin $S(\Psi) = \frac{1}{2}$; $(\gamma ^{\mu})$ is a field of Dirac matrices depending on scale gauge.
  Weyl weights are  $w(\Psi)= - \frac{3}{2}\; , w(\gamma ^{\mu })= -1$.  
 $\tilde{\phi}$ is the  scalar field (spin $0$) $w(\tilde{\phi})=-1$, extended to an isospin $\frac{1}{2}$ bundle, i.e., locally with values in $\C^2$.
$\tilde{D}_{\mu}$ denotes the covariant derivative lifted to the spinor bundle, respectively the isospinor bundle,  taking the electroweak connection with $W_{\mu \nu} $ (values in $su(2)$) and $B_{\mu \nu} $ (values in $u(1)_Y$) into account ($w(W_{\mu \nu})= w(B_{\mu \nu})= 0$)     \cite{Drechsler:Higgs}. After substitution of $D_{\mu }$ by $\tilde{D_{\mu}}$ in $L^{(\phi )}$, the total Lagrangian becomes
\beqarr
L = \frac{[c^4]}{16 \pi N} \left( L^{(HE)} +  L^{(\phi )} +    L^{(ew)}  + L^{(Y)} + L^{(\Psi )} +   \ldots \right)  
 \label{total Lagrangian} \\
\hspace*{60mm} \dots +  L^{(m)}   + [ L^{(em)}] \; .  \nonumber  
\eeqarr
The electromagnetic action  $L^{(em)} $  arises
 after symmetry reduction, induced by fixing the gauge of  electroweak symmetry imposing the condition $\tilde{\phi}_o=(0,|\phi|) $.  In $|\phi|$-gauge it is normed to a constant. {\em In this context}  Drechsler sets it to the ew energy scale, 
 $ \sqrt{2} |\phi| [\hbar c]^{-\frac{1}{2}}\doteq v \approx 246 \, GeV$, i.e., scaled down to laboratory units by a ``global''  factor $10^{-17}$ with respect to (\ref{G phi}). 

The infinitesimal operations of the ew group then lead  to a ``non-linear realization''  in the stabilizer $U(1)_{em}$ of $\tilde{\phi}_0$ and contribute to the covariant derivatives and the energy momentum tensor of the $\phi$ field.
In this way the   energy-momentum tensor of $\phi$ indicates  the {\em acquirement of mass} of the electroweak bosons,  even without assuming a ``Mexican hat'' type potential  and without any need of a  
 speculative symmetry break in the early universe.

  Drechsler's study shows that {\em mass may be acquired  by  coupling the
 ew-bosons  to gravity} through the intermediation of the $\phi$-field. This is a conceptually  convincing alternative to the usual Higgs mechanism. In similar approaches,  \citeasnoun{Hung_Cheng:1988} and \citeasnoun{Pawlowski/Raczka} have arrived at a similar expressions by deriving mass terms  perturbatively on the tree level from the same scale invariant Lagrangian without the Mexican hat potential. This is a remarkable agreement. With the Large Hadron Collider (LHC)  coming close to starting its operation,  such considerations  deserve more attention by theoretical high energy physicists.   

 In their investigation Drechsler and Tann   consider a  mass term of the scalar field as a {\em scale symmetry} breaking device, by substituting  $- M_o^2 |\phi|^2$ in $V(\phi)$ for $-\lambda _2|\phi|^2$, where $w(M_o)=0$, $M_o\neq 0$ \cite{Tann:Diss,Drechsler/Tann,Drechsler:Higgs}. This choice, although possible,  is not  compulsory  for   the  analysis of a Higgs-like mechanism which couples 
$ew$-bosons to gravity, neither does it seem advisable.
The similarity of this approach to the one  of  \cite{Pawlowski/Raczka} which relies on unbroken conformal  scale covariance and a conformally weakened gravitational action,  indicates that this type of  coupling does not depend on the scale breaking condition $w(M_o)=0$ for a  scalar field mass.  
Therefore it seems preferable 
to assume $M_0=0$  (or if  $M_0 \neq 0$,  gauge weight  $w(M_o)=-1$ and $M_o \equiv \lambda _2$), in order to keep closer to the Weyl geometric setting.

\section{Variational equations}
Variation with respect to $ \phi^{\ast}$ leads to a Klein-Gordon  equation for the scalar field  \cite[(2.13)]{Drechsler/Tann}, which  couples to scalar curvature, 
\beq   \label{K-G equation}{D}^{\mu} {D}_{\mu } \phi +  \left(    \xi \overline{R}   +  \frac{2}{\phi} \frac{\partial V}{\partial \phi^{\ast}}  \right) \phi = 0 \;  \eeq

The  factor 
\beq   \xi \overline{R}  +  \frac{2}{\phi} \frac{\partial V}{\partial \phi^{\ast}}    =: \tilde{m}_{\phi}   \; \label{m_phi}^2 =  \frac{m_{\phi}^2 c^4}{\hbar^2} \eeq
functions as a  mass-like factor of the $\phi$-field.
 The contributions of the quadratic and biquadratic terms of $V$ to (\ref{m_phi}) are intrinsic to the 
$\phi$-field. If  they vanish, $m_{\phi}$ is  derived exclusively from  
the mass-energy content of spacetime  via scalar curvature $\overline{R}$ and the Einstein equation.

Variation of flow lines with adjustment of $\rho$ such that the mass energy current is conserved, i.e.  respects the constraint (\ref{fluid constraint}),  leads to an Euler equation for the acceleration of the flow
 $\dot{X}^{\mu } := D_{\lambda }X^{\mu } X^{\lambda  }    $,
\beq   (\rho +p_m)\, D_{\lambda }X^{\mu } X^{\lambda  }   = - \partial _{\lambda } p_m \; (g^{\lambda \mu }  +  X^{\lambda  } X^{\mu } )  \; ,  \label{Euler equation}   \eeq 
where $p_m= \mu^2 \frac{d \epsilon }{d \mu }$ is the pressure of the fluid and $\rho =\mu (1+\epsilon )$ as above (\ref{rho}), cf.  \cite[96]{Hawking/Ellis} for the semi-Riemannian case. 

Variation with respect to  $\delta  g^{\mu \nu }$ gives the scale covariant Einstein equation
\beq  Ric - \frac{\overline{R}}{2} g   =   (\xi |\phi|^2)^{-1}\, \left( T^{(m)} + T^{(\phi )} + [ \ldots + T^{(Z)} \ldots ] \right) \label{Einstein equation}
    \eeq
with  a classical matter tensor compatible with (\ref{Euler equation})
\beq
T^{(m)}_{\mu \nu } = - 2 \frac{1}{\sqrt{|g|} } \frac{\delta {\mathcal L}^{m}}{\delta g^{\mu \nu }} =  (\rho + p_m)  \, X_{\mu} X_{\nu } + p_m  \, g_{\mu \nu }  \, ,\label{classicalmatter tensor}  \eeq
and field tensors 
\[
T^{(Z)} := - 2\frac{1}{\sqrt{|g|} } \frac{\delta {\mathcal L}^{(Z})}{\delta g^{\mu \nu }} \, , \quad \mbox{$Z$   for $\phi, ew,  \Psi, {Y} $}\, ,  \]
 of scale weight $w(T)=-2$. They are calculated in \cite{Drechsler:Higgs}; here we only need $T^{(\phi)}$. 

$\tilde{G}$  defined by generalization of (\ref{G phi})
\[ \frac{8\pi \tilde{G}}{[c^4]}:= \xi^{-1} |\phi|^{-2} \] may  be considered as a {\em scale covariant version of the  gravitational ``constant''}. Its weight  corresponds to what one  expects from considering the physical dimension  of $G$, 
\[ [[G]] = [[L^3 M^{-1} T^{-2}]] = [[L M^{-1}]]=2  = w (\tilde{G})\, .\] 
$L,M,T$ denote length, mass, time quantities respectively, $[[ \ldots]]$ the  corresponding metrological (``phenomenological'') scale weights. $w(\tilde G)$ correctly cancels the weight of the doubly covariant energy-momentum tensor with $w(T_{\mu \nu} )=-2$. If one wants, one may even  find {\em some} of the intentions of the original J-B-D theory (e.g., ``time dependence'' of the gravitational constant) reflected in the behaviour of the Weyl geometric $\tilde{G}$ in Riemannian gauge.  In $|\phi|$-gauge the gravitational coupling is constant; that  corresponds to the
 Einstein frame picture of conformal J-B-D theory and underpins the  importance of this gauge as a {\em good candidate}  for proportionality to measurements according to atomic clocks, without further reductions like (\ref{observable}).

The  r.h.s. of (\ref{Einstein equation})  will be abbreviated by
\beq \Theta :=   (\xi |\phi|^2)^{-1}\, \left( T^{(m)} + T^{(\phi )} + [ \ldots + T^{(Z)} \ldots ] \right) \,  \label{total Theta}\eeq 
and its constituents by $ \Theta^{(m)}  :=   (\xi |\phi|^2)^{-1}\,  T^{(m)} $ etc.

The energy-stress tensor of the   Weyl geometric $\phi$-field has been calculated by \cite[equ. (372)]{Tann:Diss} and \cite[(3.17)]{Drechsler/Tann}. It is consistent with the (non-variationally motivated) proposal of  \citeasnoun{Callan_ea} to consider an ``improved'' energy tensor: 
\beqarr  T^{(\phi)} &=&  D_{(\mu }\phi ^{\ast}  D_{\nu )}\phi  
- \xi D_{(\mu }   D_{\nu )}(\phi ^{\ast} \phi)  \nonumber \\
&-&  g_{\mu \nu } \left( \frac{1}{2} D_{\lambda }\phi ^{\ast}D^{\lambda} \phi - \xi D^{\lambda} D_{\lambda}(\phi ^{\ast} \phi) -  V((\phi ^{\ast}, \phi)
 \right)
 \label{energy stress phi}
\eeqarr

Crucial for  Drechsler/Tann's calculation is the observation that the coupling of $\overline{R}$ with $|\phi|^2$ leads to additional, in general non-vanishing, terms for the variational derivation $\delta  g^{\mu \nu }$ of the Hilbert-Einstein Lagrangian. These terms (those with factor $\xi$ in the formula above)  agree with the additional terms of the ``improved'' energy tensor of Callan-Coleman-Jackiw \cite[98--100]{Tann:Diss}.

 This modification has  to be taken into account  also in  scalar-tensor theories more broadly. Although it is being used in some of the present literature, e.g. \cite{Shojai_ea:1998},  it  has apparently found no  broad attention. In \cite[(7.29)]{Faraoni:2004} 
  the ``improved'' form of Callan energy-momentum tensor is discussed as one of  several different alternatives for an ``effective'' energy-momentum tensor. Faraoni sees here the source of the   problem of non-uniqueness of the  ``physically correct'' energy momentum tensor of a scalar field. 

Tann's and Drechsler's derivation shows, however, that there is a clear and unique variational answer  (\ref{energy stress phi}) to the question.
It   also indicates that  the  {\em truncated form of the energy tensor} (without the $\xi$-terms) {\em is in general incorrect} for theories  with a quadratic coupling of the scalar field to the Hilbert-Einstein action, independent of the wider geometrical frame (conformal semi-Riemannian or Weyl geometric). We shall see that already for  simple examples  this may have important dynamical consequences (section 9). 

We even may conjecture that  (\ref{energy stress phi}) opens a path towards a solution of  the
 long-standing problem of {\em localization of gravitational energy}, mentioned  in this context by other authors, cf. \cite[157]{Faraoni:2004}. As   $\phi$  is   an integral part  of the gravitational structure,   one may guess that (\ref{energy stress phi}) itself  may represent   the energy stress tensor of the gravitational field. At least, it is a well-defined energy tensor and is closely related to the gravitational structure. Moreover it is uniquely defined by the variational principle.

$\Theta ^{(\phi)}=\xi^{-1}|\phi|^{-2}T^{(\phi)}$ decomposes (additively) into a {\em vacuum-like}  term proportional to the Riemannian component of the metric
\beqarr    \Theta_{\mu \nu }^{(\Lambda )} &=& - \Lambda \, g_{\mu \nu } \; , \quad \mbox{with } \;\; \nonumber  \\
  \Lambda :&=&  (\xi |\phi|^2)^{-1}\, \left(  \frac{1}{2} D_{\lambda }\phi ^{\ast}D^{\lambda} \phi - \xi D^{\lambda} D_{\lambda}(\phi ^{\ast} \phi) -  V((\phi ^{\ast}, \phi)
 \right)
   , \label{Lambda}  \eeqarr
and a  {\em matter-like} residual term 
\beq   \Theta _{\mu \nu }^{(\phi \, res)} =   (\xi |\phi|^2)^{-1}\,  \left(  D_{(\mu } \phi ^{\ast} D_{\nu )}\phi 
- \xi D_{(\mu } D_{\nu )}(\phi ^{\ast} \phi) \right)  \; . \label{residual term} \eeq

Clearly $\Lambda$ is  no constant but a scale covariant quantity (of weight $-2$). 
By the scalar field equation (\ref{K-G equation}) {\em it  depends   on scalar curvature of spacetime   and  matter density}.  Its weight is $w(\Lambda )=-4$.

If $ T^{(\phi \, res)} $   reduces to  its $(0,0)$ component, it acquires the form of  a  {\em dark matter}  term. There are indications  that the scalar field contributions close to galactic mass concentrations may be helpful for understanding  dark matter  \cite{Mannheim:2005,Mielke_ea:dark_matter_halos}.

Another energy-momentum tensor of a long range field  (after $ew$ symmetry reduction) is  the e.m. energy stress  tensor. As usual it is 
\beq
T_{\mu \nu } ^{(em)} = \frac{[c]}{4 \pi}  \left(  F_{\mu \lambda } F^{\lambda}_{ \nu } -\frac{1}{4} g_{\mu \nu  } F_{\kappa \lambda }  F^{\kappa \lambda }\right) \, . \eeq
In our context $T^{(em)}$ is negligible. So is the internal energy of the fluid. For the purpose of a first idealized approximation of cosmic geometry in the following sections we work with $ \epsilon =0 $, i.e. with dust matter, 
\beq  \quad p_m=0 \, , \quad \rho = \mu  \, .  \eeq
For the sake of abbreviation we also use  the  matter density parameter
\beq \tilde{\rho } = \xi^{-1}|\phi| ^{-2}\rho   \, . \label{rho tilde}
\eeq
Remember that we have not included a dynamical term proportional to $f^{\mu \nu }f_{\mu \nu }$ into the Lagrangian, $f := d \varphi$  curvature of the Weylian scale connection.   So we exclude, for the time being,  considerations which might become crucial close to the Planck scale,  presumably supplemented by scale invariant higher order terms in the curvature \cite{Smolin:1979}, mentioned at the end of section 3.  

\section{Cosmological modeling}

Any semi-Riemannian manifold can be considered in the extended framework of integrable Weyl geometry. For  cosmological studies, Robertson-Walker manifolds are particularly important. They are  spacetimes of type
\[  M \approx  \R \times M_{\kappa }^{(3)} \]
with $M_{\kappa }^{(3)} $ a  Riemannian 3-space of constant sectional curvature $\kappa $,  usually (but not necessarily) simply connected. If in spherical coordinates $(r,\Theta, \Phi )$ 
\beq d \sigma _{\kappa } ^2  =  \frac{dr^2}{1- \kappa r^2} +r^2 (d\Theta ^2  + \sin^2 \Theta \, d \Phi^2) \label{metric constant curvature} \eeq 
 denotes the metric on  the spacelike fibre $M_{\kappa }^{(3)} $, the Weylian metric $[(g, \varphi )]$ on $M$ is specified by its Riemann gauge $(\tilde{g},0)$ like in standard cosmology:  
\beq \tilde{g}: \quad \quad ds^2 = - d{\tau }^2 + a(\tau )^2 d \sigma _{\kappa } ^2 \; \label{Robertson Walker metric}\eeq 
 $\tau = x_0 $ is a local or global coordinate (cosmological time parameter of the semi-Riemannian gauge) in $\R$, the first factor  of $M$.  We shall speak of 
{\em Robertson-Walker-Weyl (R-W-W) manifolds}.

In the semi-Riemannian perspective 
$a(\tau )$, the {\em warp function} of $(M, [(g,\varphi)])$, is usually interpreted as an expansion of space sections.  The Weyl geometric perspective shows that this need not be so.
For example,  there is a gauge $(g_w, \varphi_w)$   in which the ``expansion  is  scaled away'': 
\beq 
g_w = \Omega _w^2 g \quad \mbox{with } \;\;  \Omega _w := \frac{1}{a} \; 
\eeq
With 
\beq  t:= \int ^{\tau } \frac{du}{a(u)} = h^{-1}(\tau ) \quad \mbox{and its inverse function} \;\; h(t) = \tau \label{warp gauge reparametrization}
\eeq
 we   get a gauge 
\beqarr g_w(x): &=& -dt^2 + d\sigma _{\kappa }^2 \;,   \label{Hubble gauge}\\
  \varphi_w (x) &=&   d \log (a \circ h)  = (a'\circ h) \, dt = a'(h(t))\, dt \; , \nonumber  \eeqarr  
in which the Riemannian component of the metric looks static. According to (\ref{f gauge}), it may be called  {\em warp gauge} of the Robertson-Walker manifold, because the warp function is scaled to a constant. The other way round, the warp function is nothing but the integrated scale transfer of  warp gauge  (\ref{scale transfer})
\beq a(p) = a(p_o) e^{\int_o^1  \varphi_w(\gamma')} \, . \label{warp}
\eeq

The geodesic path structure is invariant under scale transformations of IWG. In agreement with Diracian geodesics of weight $-1$ the  observer field  $X^i = \partial$/$ \partial x_i$ has also to be given the  weight $w(X)=-1$. Then the energy $e(p)$ of a photon along a null-geodesic $\gamma $, observed at $p$ by an observer of the family $X$ is given by
\[ e(p) = g_p(\gamma '(p), X(p))  \, .\]
It is of weight $w(e) = 2-1-1=0$ and thus  scale invariant. Therefore {\em redshift}  (cosmological or gravitational) of a photon emitted at $p_o$ and observed at $p_{1}$ with respect to observers of the family $X$, 
\beq z+1 = \frac{e(p_o)}{e(p_{1})} = \frac{g_{p_o}(\gamma '(p_o),X(p_o)) }{g_{p_1}(\gamma '(p_1),X(p_1))} \, , \label{z+1 basic}
\eeq
 is also {\em scale invariant} ($\gamma $ null-geodesic connecting $p_o, p_1$).

We see that  cosmological redshift is not necessarily characterized by a warp function $a(x_0)$;  in warp gauge it  is expressed  by the scale connection $\varphi_w$ 
and can be  read off directly from Weyl's ``length'' transfer because of $z+1 = \frac{a(p)}{a(p_o)}$ and (\ref{warp}):
\beq z +1 = e^{\int_0^1 \varphi (\gamma ') d\tau } = \lambda (p_o,p_1)\; ,   \quad \gamma \; \mbox{connecting path. } \;\; \label{z+1} \eeq
 We therefore call $ \varphi_w$ the {\em Hubble connection} of the R-W-W model. It is timelike, $\varphi_w = H(t) dt$, with $H(t)=a'(h(t))$.

If Hubble redshift is not due to  space expansion but to a field theoretic energy loss of photon energy with respect to the observer family, the warp gauge picture will be more appropriate to express physical geometry than Riemann gauge. In this case, the Hubble connection should not be understood as an independent property of cosmic spacetime, but rather 
 depends on the mean mass-energy density in the universe. Different authors starting from \cite{Zwicky:Redshift} to the present have tried to find a higher order gravitational effect which establishes such a relation. A convincing answer has not yet been found.  If however the Hubble connection is ``physical'', {\em Mach's principle} suggests that  it should be due to the mean distribution of cosmic masses.  As   simplest possibility, we  may conjecture that  a linear relation between $H^2$ and mass density might hold in large means {\em in warp gauge}, 
\beq    H^2 =  \eta _1 \tilde{\rho} + \eta _0\, , \quad  \eta_1 > 0  \quad \quad \mbox{\em ($H^2$ conjecture). } \;\; \label{H^2 conjecture}
\eeq
In the models studied below such a coupling of $H^2$ to mass density is  a consequence of the scalar field equation and the Einstein equation (\ref{H^2 Weyl universe}).

As geodesic distance  (\ref{geodesic distance}) is no local observable and not scale invariant,  the question arises   which of the gauges, Riemann or  warp  gauge gauge (or any other one), expresses the measurement by atomic clocks.  In the context of Weyl geometric scalar field theory the question can be reformulated: Does $|\phi|$-gauge coincide with one of these gauges and if so, with which? Ontologically speaking, the two gauges, Riemann or warp, correspond to {\em two different  hypotheses} on the cause of cosmological redshift: {\em space expansion} (Riemann gauge) or a {\em field theoretic energy loss of photons} (warp gauge). Weyl geometry allows to translate between the two  hypotheses and provides a theoretical framework  for a systematic comparison.

\section{Weyl universes}
In order to get a feeling for the new perspectives opened up by Weyl geometric scalar fields, we investigate  the simplest examples of R-W-W cosmologies with redshift. 
In warp gauge their Weylian metric is given by a constant Riemannian component $ ds^2 =  - dt^2 + d\sigma^2_{\kappa}$   (where  $\kappa \in \R $ denotes the sectional curvature of the spatial fibres $M_{\kappa }^{(3)} $) and a constant scale connection with only a time component:
\beqarr ds^2 &=&    - d t^2 + \left(\frac{dr^2}{1-\kappa\, r^2} + r^2 d\vartheta ^2 +  r^2 \sin^2\vartheta \, d\phi^2 \right) \label{Weyl universe} \\
 \varphi &=& (H, 0,0,0)\, , \quad H > 0\quad \mbox{constant } \nonumber
 \eeqarr 
$H$ is called  Hubble constant (literally). 
We encounter here a Weyl geometric generalization of the {\em classical static models} of cosmology, but now {\em including redshift} (\ref{z+1}):  
\[ z+1 = e^{H (t_2-t_1)} \]
These models will be called {\em Weyl universes}. Different to the classical static cosmologies  this type of static geometry can be upheld, under certain conditions, in a natural way by the  dynamical effects of the scalar field $\phi$, respectively the Weylian scale connection $\varphi$ (section 9).

The integration of the scale connection leads to an exponential length transfer function $\lambda (t) = e^{Ht}$. Transition to the  R-W metric presupposes a change of the cosmological time parameter $\tau = H^{-1}e^{Ht}$; then
\beq   a(\tau) = H \tau \, .   \label{Weyl universe R-W} \eeq 
 Thus this class deals with a Weyl geometric version of {\em linearly warped} (``expanding'') R-W cosmologies.

Up to (Weyl geometric) isomorphism, Weyl universes are characterized by {\em one} metrical parameter ({\em module}) only, 
\beq \zeta := \frac{\kappa}{H^2} \, .  \label{zeta}\eeq

In warp gauge the components of the affine connection with respect to spherical coordinates (\ref{Weyl universe}) are
\beqarr \Gamma _{00}^0 &\doteq & H \, , \hspace{25mm} \Gamma _{0 \alpha }^{\, \alpha } \doteq H \quad \quad (\alpha = 1,2,3) \label{L-C connection Weyl universe}\\
\Gamma _{11}^0 &\doteq& H (1-\kappa r^2)^{-1} \; , \quad 
  \Gamma _{22}^0  \doteq H r^ 2 \;  ,\quad   \Gamma _{33}^0  \doteq H r^ 2 \sin^2 \vartheta \nonumber \\
\Gamma _{\alpha \beta }^{\gamma } &\doteq&  {}_g\Gamma _{\alpha \beta }^{\gamma } \quad (\alpha , \beta , \gamma  = 1,2,3) \nonumber           
\eeqarr
Ricci and scalar curvature are
\beqarr Ric &\doteq& 2 (\kappa +H^2) d \sigma _{\kappa }^2 \label{curvature Weyl universe} \\
\overline{R} &\doteq& 6 (\kappa + H^2) \; . \nonumber
\eeqarr
Similar to those of the classical static models, Weyl universes have constant entries of the energy momentum tensor but contain quadratic Hubble terms $H^2$ in addition to spacelike sectional curvature terms: 
\beq  \Theta _{00} \doteq 3 (\kappa +H^2) \, , \quad \quad  \Theta _{\alpha \alpha } \doteq -(\kappa +H^2) (d \sigma^2 _{\kappa })_{\alpha \alpha }  \quad  \quad  (\alpha  = 1,2,3) \label{energy-momentum Weyl universe}\eeq 
That corresponds to a total energy density $\rho $ and  pressure $p$
\[ \tilde{\rho } \doteq 8 \pi G \, \rho \doteq  3 (\kappa +H^2)\, , \quad \quad \quad \tilde{p} \doteq 8 \pi G \, p \doteq  - (\kappa +H^2) \]
For $\kappa >0$ we obtain {\em Einstein-Weyl models} similar to the classical Einstein universe. 

The  next  question will be, whether the equilibrium condition between energy density and negative pressure  necessary for upholding such a geometry may be  secured by the scalar field.

\section{Energy momentum of the scalar field and dynamical consistency}
Here and in the following sections we work in warp  gauge, i.e., spacelike fibres $M_{\kappa }^{(3)} $ are gauged to constant (time-independent) sectional curvature,  if not stated   otherwise. Therefore the following equations are in general no longer scale invariant. 

In warp gauge 
 the Beltrami-d'Alembert operator of Weyl universes is given by
\beq  \boxempty \phi = D^{\lambda }D_{\lambda } \phi \doteq -(\partial _0^2 - 3 H\partial _0 + 2 H^2) \phi +    \tilde{\nabla }^{\alpha }\tilde{\nabla }   \phi \, , \eeq 
where $\lambda = 0, 1, 2,3 \,, \;  \alpha = 1, 2, 3$, and 
 $ \tilde{\nabla }^{\alpha }$ denotes the covariant derivatives of the Riemannian component of the metric  {\em along spatial fibres}. 

Separation of variables 
\[ \phi (t, \tilde{x})= e^{i \,\omega t} f(\tilde{x})\; , \quad \quad \tilde{x}= (x_1, x_2, x_3)\, , \; \omega \in \R\, , \]
with an eigensolution $f$ of the Beltrami-Laplace operator and $w(f)=-2$,
\[    \tilde{\nabla }^{\alpha }\tilde{\nabla }_{\alpha } f = \lambda \,  f  \; , \quad \quad \lambda \in \R \,  , \]
leads to
\[
\boxempty  \phi \doteq  (\omega ^2 +3 H^2 \omega \, i + \lambda -2H^2) \phi \, .  \]
Reality of  the mass like factor of the K-G equation implies $ \omega = 0$.
Moreover, for Einstein-Weyl universes, $\kappa > 0$, $f$ is a spherical harmonic on the 
3-sphere $S^3$ with eigenvalue $\lambda $. The only spherical harmonic with  constant norm is  $f \equiv const$, $\lambda =0$, and thus  
$\phi = {\mathcal Re} (\phi)= const$ is a ground state solution (after separation of variables) of the scalar field equation  in warp gauge. 
We conclude that  {\em warp gauge of Weyl universes   coincides}  {\em with $|\phi|$-gauge} and 
\[  \boxempty  \phi \doteq   - 2 H^2 \phi \, . \].  
As we work in this section in $|\phi|$ gauge anyhow, the denotation $\phi_{\ast}$ used  where different gauges are compared  is here simplified to $\phi$.  

Using (\ref{curvature Weyl universe}) we see that the K-G equation  (\ref{K-G equation})  is satisfied, iff
\beq  H^2 \doteq   \kappa  +  \frac{2}{\phi}\frac{\partial V}{\partial \phi^{\ast}}  \, . \label{K-G Weyl universe}
\eeq

The contributions to the energy momentum tensor   (\ref{energy stress phi})  in $|\phi|$-gauge (identical to warp gauge) are given by:
\[  D_0 \phi  \doteq  -H \phi \, , \quad D_{\alpha }  \phi \doteq 0 \quad  (\alpha =1,2,3) \]
and therefore
\beq  D_{\lambda}  
\phi^{\ast} D^{\lambda} \phi \doteq - H^2 |\phi|^2   \, . \eeq

For  constant functions $f$ of weight $w(f)= -2$, like $ |\phi|^2 $, the scale covariant gradient   {\em does not vanish} because of the weight correction of (\ref{scale covariant derivative}), 
\[  D_0 f \doteq (\partial _0 + w(f) H) f \doteq - 2H f \, , \quad \quad D^0 f \doteq 2H f \, , \]
while all other components vanish,  $D_{\alpha }f \doteq 0$. For the next covariant derivative the only non-vanishing component of the affine connection is $\Gamma ^0_{00} \doteq H$. That leads to
\[  D_0 D_0 f \doteq (\partial _0  - \Gamma ^0_{00}  + w(D_0 f)\, H )\,D_0 f \doteq 6 H^2 f \, . \]
Thus $\frac{1}{2} D_0\phi^{\ast}D^0 \phi - \frac{1}{6} D^0D_0 |\phi|^2 \doteq \frac{H^2}{2}|\phi|^2$. From  (\ref{Lambda})  we find  
\beq \Lambda  \doteq   3H^2  - 6 \frac{V}{|\phi|^2} \; .\label{Lambda Weyl universe}  \eeq
The truncated version of  (\ref{energy stress phi}) would lead to a different value and thus to a completely different dynamics of the whole system.

 $T^{(\phi \, res)}$ vanishes
and 
\beq \Theta^{(\phi)}  = - \Lambda \, g  \; . \label{energy stress phi Weyl universe}
 \eeq
Formally $\Theta^{(\phi)}$  looks like the ``vacuum tensor'' of the received approach. Note, however, that here the  coefficient $\Lambda $ is no universal  constant but couples to the mass content of the universe via $H^2$, the relation  (\ref{K-G Weyl universe}) and the   energy  component of the  Einstein equation (\ref{Einstein equation}). The $(0,0)$-component of the latter is 
\beqarr 3 (\kappa + H^2 ) &\doteq& \tilde{\rho }  + \Lambda \doteq \tilde{\rho }  +3 H^2 - 6 \frac{V}{|\phi|^2} \, ,   \nonumber \\
3 \kappa   &\doteq&    \tilde{\rho }  - 6 \frac{V}{|\phi|^2} \, ,\label{Einstein equ 00}
\eeqarr
where the convention (\ref{rho tilde}) for $\tilde{\rho}$ has been applied.  (\ref{K-G Weyl universe})   and the observation that in the $V$ considered here $ \frac{1}{\phi}  \frac{\partial V}{\partial \phi^{\ast}}- \frac{V}{|\phi|^2}  = \lambda _4 |\phi|^2$ imply
\beq H^2 \doteq \frac{\tilde{\rho}}{3} +2 \lambda _4 |\phi|^2 , \quad \quad
 \Lambda \doteq   \tilde{\rho }  - 6 \lambda _2 \, .  \label{H^2 Weyl universe}\eeq 
Thus  the  $H^2$-conjecture (\ref{H^2 conjecture}) turns out to be true for the case of  Weyl universes. In agreement with Mach's ``principle''  $H^2$ and $\Lambda$   depend on the mass content of the universe. This  agrees with the 
  basic principle of physics  that a causally important  structure of spacetime and matter dynamics should not be independent of the matter content of the universe. The basic principle  is satisfied for the energy tensor of the scalar field of Weyl universes and for $\Lambda$  in general. According to  \cite{Fahr:BindingEnergy} 
this is  a {\em desideratum for any realistic cosmological model}.

Now it is clear  which conditions have to be satisfied by $\phi$ and $V(\phi)$,  if  Weyl universes are to be kept in  equilibrium by the scalar field. In this case the total amount of mass-energy density and the negative pressure of the
 vacuum-like term of the scalar field tensor have to counterbalance each other ({\em dynamical consistency} of Weyl universes). 
The general balance condition for total energy density $\rho $ and  pressure $p$ of fluids  is
 \[ \rho +3p = 0 \, .\]
Because of $\tilde{p} = - \Lambda $ it becomes in Einstein universes:  
\beqarr  \tilde{\rho } + \Lambda \doteq \Theta _{00} &\doteq& 3 \Lambda \quad  \longleftrightarrow \nonumber \\
   \tilde{\rho} &\doteq& 2  \Lambda  \, \quad \quad \quad  \label{Hyle}\eeqarr

Altogether  K-G equation (\ref{K-G Weyl universe}), Einstein equation (\ref{Einstein equ 00}) and the consistency condition for Weyl universes (\ref{Hyle}), including (\ref{Lambda Weyl universe}),  give an easily surveyable set of conditions (the Euler equation (\ref{Euler equation}) is trivially satisfied)  
\beqarr 
\kappa + 2 \lambda _2 + 4 \lambda _4 |\phi|^2 &\doteq& H^2  \nonumber \\
3 \kappa    -  \tilde{\rho }  + 6 \lambda _2 + 6 \lambda _4 |\phi|^2 &\doteq& 0     \label{system}\\
  \tilde{\rho} +  12 \lambda _2 + 12 \lambda _4 |\phi|^2  &\doteq& 6 H^2 \nonumber  
\eeqarr 
To get a first impression what this means in terms of energy densities for low values of $\zeta $ we list some examples including comparison with  $\rho _{crit}= \frac{3H^2[c^4]}{8 \pi G}\doteq 3H^2 \xi |\phi_{\ast}|^2$  \\[0.5em]
\noindent
{\em Examples:} A moderate curvature module $1$ arises for
   \beqarr 
   \tilde{\rho} \doteq 4 H^2 \, , \quad \Lambda &\doteq&  2 H^2 \, , \quad \kappa \doteq H^2 \,  
  \quad \lambda _2 \doteq  \frac{H^2}{3}  \, ,\quad \lambda _4 |\phi|^2 \doteq - \frac{H^2}{6} \, ,   \nonumber \\
   \Omega _m  &=& \frac{4}{3}, \quad  \Omega _{\Lambda } = \frac{2}{3}\; , \quad   \zeta = 1 \, .\quad \quad 
\label{example 1} 
 \eeqarr 
For $\lambda _4=0$, on the other hand, we get 
\beq \Omega _m  = 1, \; \Omega _{\Lambda } = \frac{1}{2}\; , \quad   \zeta =   \frac{1}{2} \, , \quad \quad \lambda _2  \doteq  \frac{H^2}{4}\,.  \quad \quad  \label{example 2}
\eeq 
It cannot come as a surprise  that we find  relatively high mass energy densities, as we are working with positive space curvature. They increase with higher curvature values, e.g., $\Omega _m  = 2, \; \Omega _{\Lambda } = 1 $ for $\zeta =2$ etc. 

Considering present mass density estimations,  this might appear as a reason to discard these models. But there are other reasons to shed a second glance at  them,  at least as  ``toy'' (i.e. methodological) constellations. In the light of the conjecture (section 6)  that  $T^{(\phi)}$ may be considered as   energy tensor of the gravitational structure  (extended by $\phi$),  these simple models demonstrate the {\em possibility} of a cosmic geometry balanced by the gravitational stress energy tensor itself (cf. in a different context \cite{Fahr:BindingEnergy}).  This may be important for attacking the  problem of stability. We do not do this here; but turn to a second glance at the empirical properties of these models. This also illuminates the more general question how  Weyl geometric models behave under empirical scrutiny.

\section{First comparison with data}
First of all, it is clear  that the precision of the empirical tests of GRT inside the solar system lie far away from cosmological corrections in any approach.  In Weyl geometry the cosmological corrections to weak field low velocity orbits amount to an additional coordinate acceleration $ \ddot x = - H\dot x =:a_H$ \cite[app. II]{Scholz:ModelBuilding}. For typical low velocities of planets or satellites $\sim 10 \; km\, s^{-1}$, this is  9 orders of  magnitude below solar gravitational acceleration at  distance $10\; AU$ (astronomical units) from the sun, and  4 orders of magnitude below the anomalous acceleration  $a_P$ of the Pioneer spacecrafts determined in the late 1990-s \cite{Anderson_ea:Pioneer_I}. 
 Present solar system tests of GRT work at an error margin corresponding to acceleration sensitivity several orders of magnitude larger   \cite{Will:LivingReviews}.

Thus, for the time being,  the Weyl geometric cosmological corrections   cannot be checked empirically  by their {\em dynamical}  effects  on the level of solar system in terms of  parametrized postnewtonian gravity (PPN). On the other hand the  Hubble connection   leads  to an additional redshift $\Delta \nu \approx H c^{-1} v  \Delta t$ over time intervals $\Delta t$ for space probes of the Pioneer type with nearly radial velocity $v$. This corresponds to the absolute value of the anomalous Pioneer acceleration, but is of  wrong sign, if  compared  with the  interpretation of the Pioneer team. Follow up experiments will be able to clarify the situation \cite{Odyssee}.

At present a  first test of the model with data from observational cosmology is possible by confronting it with the  high precision supernovae data available now for about a decade \cite{Perlmutter:SNIa}, recently updated \cite{Riess_ea:2007}. In  the Weyl geometric  approach the  damping of the energy flux of cosmological sources  is due to four independent contributions: In addition to    damping by
 redshift $\sim (1+z)$ (energy transfer of single photons with respect to $X$), the internal  time dilation due to scale transfer of time intervals (\ref{scale transfer}) reduces the flux  by another factor $\sim (1+z)$ (reduction of number of photons per time). Moreover the area increase $A(z)$ of light spheres at redshift $z$ in the respective geometry (here in spherical geometry)  and an  extinction exponent $\epsilon $ have to be taken into account.  As distance   $d \sim (1+z)$,  the  absorption contributes another factor  $\sim (1+z)^{\epsilon }$. The energy flux $F(z)$ is thus given by 
\[  F(z)  \sim (z+1)^{-(2+ \epsilon )}   A(z) ^{-1}     \]
For the module $\zeta = \kappa H^{-2}$  the area of spheres is
\[ A(z) = \frac{4\pi }{\kappa } \sin ^2 (\sqrt{\zeta }\ln(1+z) ) \, . \]
Then the logarithmic relative magnitudes $m$ of sources with absolute magnitude $M$  become (cf. \cite{Scholz:ModelBuilding})
\beq  m(z,\zeta,\epsilon ,M)= 5\log_{10} \left[  (1+z)^{\frac{2+\epsilon }{2}} \zeta^{-\frac{1}{2}} \sin \left( \log (1+z) \, \zeta^{\frac{1}{2}}  \right) \right] + C_M \eeq
where the constant $C_M$ is related to the absolute magnitude of the source by  
\[ C_M = M - 5 \log_{10}(H_1 10^{-5}) \; .  \]

A fit of the redshift-magnitude characteristic of Einstein-Weyl universes with the set of 191 SNIa data  in \cite{Riess_ea:2007} leads to best values  $ \epsilon_0 \approx 1 $  and 
$ \zeta_0 \approx 2.5$ 
with  confidence intervals $1.46 \leq \zeta \leq 3.6$ and  $0.65 \leq \epsilon \leq 1.3$.\footnote{In \cite[equ. (38)]{Scholz:ModelBuilding}   a   more aprioristic deduction of the energy damping has been used, fixing $\epsilon$ to $1$. The  empirically minded approach given here follows a proposal of  E. Fischer (personal communication).}
The root mean square error is
$\sigma_{Weyl\,} (\zeta )  \leq 0.22 $ and increases very slowly with change of $\zeta$. In the whole confidence interval it   is  below the mean square error of the data  $\sigma_{dat} \approx 0.24$ (given by Riess e.a.) and below the error of the standard model fit $\sigma_{SMC} \approx 0.23$. For $\epsilon \approx 0.65$ (lower bound of its confidence interval) the root mean square error of the Weyl model predictions for $\zeta = 0.5$, compared with the data, is still $\sigma_{Weyl\,} (\zeta ) \approx  0.231$ comparable to the quality of the SMC and $ < \sigma_{dat}$. It surpasses the latter only for $\zeta \leq 0.2 $. According to this criterion our examples (\ref{example 1}), (\ref{example 2})   survive the test of the supernovae data as well as the standard approach.

\begin{figure}[h]
\center{\includegraphics*[scale=1.2]{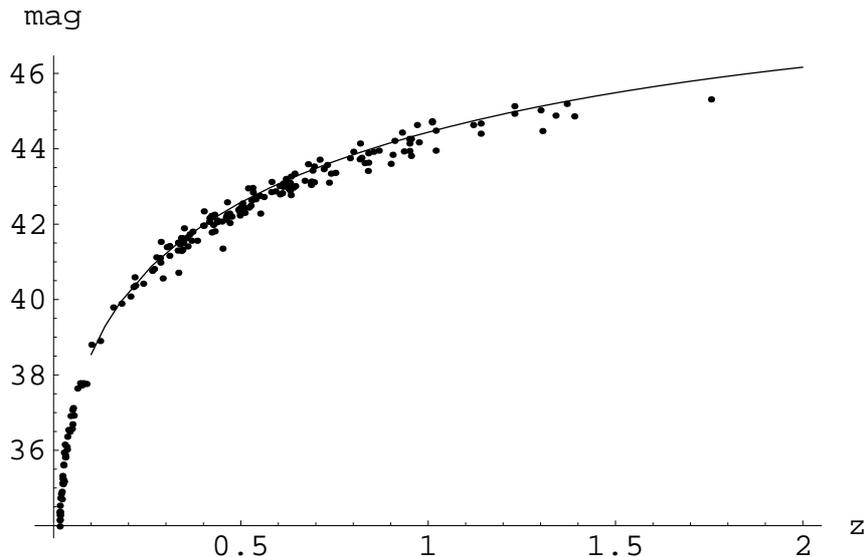}}
\caption{Magnitudes of 191 supernovae Ia (mag) $z\leq 1.755 $  (Riess e.a. 2007),
  and prediction in Weyl geometric model $\zeta =1$} 
\end{figure}

At the moment supernovae data do not  allow to  discriminate between the Weyl geometric approach and the Friedman-Lemaitre one of the SMC. That may change, once precise supernovae data are available in the redshift interval $2 < z < 4$. 

Of course many more data sets have to be evaluated, before a judgment on the   empirical reliability of the Weyl geometric approach can be given.  The cosmic  microwave background, e.g., appears in our framework  as a thermalized background equilibrium state of the quantized Maxwell field. A corresponding mathematical proof of a perfect Planck spectrum of  a high entropy state of the Maxwell field in the Einstein universe has  been given by \cite{Segal:CMB1}. Anisotropies seem to correlate with inhomogeneities of nearby mass distributions in the observable cosmological sky by the Sunyaev-Zeldovic (SZ) effect \cite{Shanks:WMAP}. For more distant clusters that is completely different: The almost lack of SZ effects for larger distances has been characterized  as ``paradoxical'' by leading astronomers \cite{Shanks/Bielby:2007}. It seems to indicate that the origin of the microwave background does not lie beyond these clusters.
 Empirically the assumption of a {\em deep redshift origin} of the anisotropies is therefore no longer  as safe as originally assumed, perhaps even doubtful.

Some empirical evidence, like quasar data,   goes similarly  well in hand with the Weyl geometric cosmological approach as with the SMC, other  worse. Most importantly, present estimations of mass density lie far below the values indicated by our models.
But the last word on mass density values may not have been spoken yet. The determination of the present values for $\Omega _m (\approx 0.25$) is strongly dependent on the standard approach of cosmology. We should not be faulted by what philosophers call the experimenters ``regress'' (testing theories by evaluatoric means which presuppose already parts of the theory to be tested) and keep our eyes open for future developments \cite{Lieu:2007,Fischer:Balance}.

On the other hand, other data are better reconcilable with the Weyl models than with the SMC. In particular the lack of a positive correlation of the metallicity  of galaxies and quasars with cosmological redshift $z$ seems  no good token  for a universe in global and longtime evolution. Moreover, the observation of high redshift X-ray quasars with very high metallicity (BAL quasar APM 08279+5255 with $z \approx 3.91$ and Fe/O ratio of about 3) appears discomforting from the expanding space perspective. Present understanding of metallicity breeding  indicates that a time interval of about 3 Gyr is needed to produce this abundance ratio, while the age of SMC at $z \approx 3.91$ is about $t \approx 1.7$ Gyr, just above  half the age needed  \cite{Hasinger/Komossa}. 

Many more data sets have to be investigated carefully comparing different points of view afforded by differing theoretical frames. It is too early to claim anything like secure  judgment on this issue.

\section{Conclusion and discussion of open questions}
The extension of the Weyl geometric approach from field theory to cosmology  leads to a formally satisfying weak generalization of the Einstein equation by making all its constituents scale covariant, equ. (\ref{Einstein equation}). The corresponding Lagrangian (\ref{Lagrangian phi}) uses a minimal  modification of the classical Lagrangians. It is inspired by a corresponding scale covariant approach to semi-classical field theory of W. Drechsler and H. Tann   and is similar to the one used in J-B-D type scalar-tensor theories.
Weyl geometric gravity theory  has features  similar to those of conformal
 J-B-D theories  (section 4); but it  builds upon different geometrical concepts.
The  scale connection $\varphi$, the specific new geometrical structure of Weyl geometry, shows remarkable physical properties. 
Integrated it describes transfer properties of metric dependent quantities (\ref{scale transfer}), and its dynamics is basically that of the scalar field. Both   can be transformed into another 
(\ref{sigma}), (\ref{varphi_ast}).
Local observables can be formulated scale gauge  invariantly (\ref{observable}), but have a direct representation in a preferred gauge  (\ref{X phi gauge}). 

A difference to large parts of the work in  J-B-D scalar-tensor theory arises from the actual consequences drawn from  coupling  the scalar field to the Riemann-Einstein action for variation with respect to $\delta g^{\mu \nu}$. Tann and Drechsler have shown that a correct evaluation of $\frac{\delta (|\phi|^2 \overline{R})}{\delta g^{\mu \nu}}$ leads to the same additional terms in the energy tensor of the scalar field (\ref{energy stress phi}) as   postulated  by Callan/Coleman/Jackiw by  different (quantum physical) considerations.  This argument has apparently found not much  attention in the literature on J-B-D theories, although it should have done so. As long as this is not the case, the dynamics of scalar fields in J-B-D theories and in Weyl geometry  seems to be different.

The scale covariant perspective  sheds new light on the class of  Robertson-Walker solutions of the Einstein equation. Weyl geometry suggests to consider non-expanding versions of homogeneous and isotropic cosmological geometries, in which the redshift is encoded by a Weylian scale connection with only a time component $\varphi = H dt$, the  Hubble connection (\ref{Hubble gauge}).  Thus the  question arises, whether the warp function of Robertson Walker models does describe a real expansion, as usually assumed, or whether it is no more than a  mathematical feature of the Riemannian component of a scale gauge  without immediate  physical significance. 

For a first approach to this question we have investigated special  solutions of the coupled system of a scale covariant Euler type fluid equation  (\ref{Euler equation}), in the simplest case dust, the scalar field equation (\ref{K-G equation}),  and a scale invariant version of the Einstein equation (\ref{Einstein equation}). This  leads to the intriguingly simple geometrical structure of   {\em Weyl universes}  (\ref{Weyl universe}) and gives a first impression of the new features which can  arise in Weyl geometric gravity. 
The  Riemannian component of the metric of these models coincides with that of  the classical static solutions of cosmology; but in addition we have a time-homogeneous Hubble connection.  
The scalar field's energy-stress tensor can be evaluated explicitly, (\ref{Lambda Weyl universe}) and (\ref{energy stress phi Weyl universe}). It shows good physical properties, {\em  if it  is considered in the untruncated form} of (\ref{energy stress phi}).   Formally it  looks  like the vacuum tensor,  $ \Theta  = - \Lambda \, g   $, of the standard approach; physically, however, it is  different. For the case of Weyl universes a close link between the  coefficient $H$ of the  Hubble connection  and mean cosmic mass energy density can be established (\ref{H^2 Weyl universe}).  This can be considered as a kind of implementation of Mach's principle.

In several aspects  Weyl  geometric models behave differently from what is known and expected for classical 
F-L models of cosmology. In this sense they  may be useful to challenge some deeply entrenched convictions of the received view: 
\begin{itemize}
\item[(1)] Cosmological redshift  need not necessarily be due to an ``expansion'' of spacetime, corresponding   to a  realistic interpretation of the warp function of Robertson-Walker models. It may just as well be expressed by the scale connection ({\em Hubble connection}) corresponding to a  Weyl geometric scalar field (\ref{z+1}), (\ref{varphi_ast}).
\item[(2)] Scalar fields can develop a dynamical contribution   (\ref{energy stress phi}) which may stabilize the geometry even to the extreme case of a  ``static'' Weyl geometry (although linearly expanding in the Riemann picture) (\ref{Hyle}). 
The arising problems of stability and of a possible tuning of the parameters $\lambda _2, \lambda _4$ of $V$ have been left open here. 
More sophisticated  examples will have to be investigated; some of them may show oscillatory behaviour. 
\item[(3)] This approach leads to cosmological models beyond the standard approach which are, in any case, methodologically interesting (\ref{example 1}), (\ref{example 2}). Perhaps they even are of wider empirical interest.
\item[(4)] Our  case study of Weyl universes shows that in particular the consequences of Tann's and Drechsler's calculation of the variational consequences of  coupling the  scalar field to the Einstein-Hilbert action have to be taken seriously already on the semi-classical level. 
\end{itemize}

The Einstein-Weyl models  should be studied more broadly 
from the point of view of observational cosmology.   Here we had to concentrate on one aspect only, the supernovae data. They are  well fitted by this models  and clearly favour positive curvature values in the Weyl universe class, $0.2 \leq \zeta < 3.6$. This indicates higher mass density values than accepted at present. Our main example (\ref{example 1}) has values 
 $\Omega _m = \frac{4}{3}, \Omega _{\Lambda }= \frac{2}{3}, \; \zeta=1$.
Many will consider this already as an indicator of lacking empirical adequacy. But we have  reasons to be more cautious in this respect; we should wait for further clarification of this issue (section 10).

Moreover, there may arise  motivations from another side 
 to improve the  approach to cosmology presented here, if the Weyl geometric approach turns out fruitful in the field theoretic sector.
The scale covariant scalar field prepares the path towards  a different approach to the usual Higgs mechanism for understanding the mass acquirement of  ew bosons \cite{Drechsler:Higgs}. An  analysis of the consequences of the Weyl geometric ``pseudo-Higgs'' $\phi$-field without  boson  for the  perturbative calculations of the  (modified) standard model of elementary particle physics is a desideratum. For a comparison with upcoming experimental results at the LHC  it may even become  indispensable. Should the empirical evidence fail to  support the present expectation of a massive Higgs boson, and even exclude it  in the whole energy interval which at the  moment is still theoretically and experimentally consistent with the present standard model EP ($120 - 800\, GeV$) after several years of data collection, the scale covariant scalar field would be a good candidate for  providing  a new conceptual bridge between elementary particle physics and cosmology. In this case the missing link  between gravity and the quantum had 
 be sought for in a  direction  explored in first aspects from the point of view of in quantum gravity, e.g., by  \cite{Smolin:1979}. If the ``pseudo-Higgs'' explanation of ew boson mass would be supported by a negative experimental results for the massive Higgs boson, we even had to face the  possibility that  effects of gravity  may be   observable in high energy physics (and in fact have been observed already)  at  laboratory energy scales,  much  lower than usually expected. 

Such strong perspectives will  very likely be  turned down or corroborated in the coming few years by experiment. Even if they should be invalidated, our cosmological considerations will not have been in vain.  The   Weyl geometric models show a route how the anomalous behaviour of cosmic vacuum energy  may be dissolved  in a weak extension of classical GRT, without sacrificing the empirical phenomena or even the overall link to experiment.  If this achievement stabilizes and can be extended to the problem of dark matter, the cosmological implications alone would  be worth the trouble to work out more details of  the Weyl geometric approach. 

\small 
\section*{Acknowledgments} An anonymous referee pushed me hard  to clarify the relationship between scale invariant features of Weyl geometry and the preferred scale gauge. F. Hehl (K\"oln) gave me the advice to analyze the relationship between the Weyl geometric approach and the scalar-tensor theories of J-B-D   (the same hint had come  from  G. Ellis, Cape Town, already years ago).  To W. Drechsler 
 (M\"unchen) I owe  the intriguing analysis referred to  in this paper several times and  also a pleasant and  intense communication about questions pertaining to Weyl geometry and the physical interpretation of the  scalar field. Discussions with H.-J. Fahr (Bonn) and E. Fischer (Stolberg)  have  enriched my understanding of the empirical context of these studies.

 Matthias Kreck (Bonn) has accompanied my path into this field, since the very beginnings, with friendly interest and support from the mathematician's side; another mathematician's talk  encouraged me to start this route of investigation \cite{Cartier:Cosmology}.
Last not least,  my colleagues from the Interdisciplinary Centre for Science and Technology Studies at the University Wuppertal  supported me, and continue to do so, with their sympathy and interest for open horizons in science.

\small
\bibliographystyle{apsr}

\end{document}